\documentclass[doublecol]{epl2} 

\title{Characterization of the Quantized Hall Insulator Phase in the Quantum Critical Regime}
\shorttitle{Characterization of the Quantized Hall Insulator Phase} 

\author{Juntao Song\inst{1,2} \and Emil Prodan\inst{1}}
\shortauthor{Song \& Prodan}

\institute{                    
  \inst{1} Department of Physics, Yeshiva University - New York, NY 10016, USA\\
  \inst{2} Department of Physics and Hebei Advanced Thin Film  Laboratory, Hebei Normal University - Hebei 050024, China
}
\pacs{nn.mm.xx}{72.25.-b, 72.10.Fk, 73.20.Jc, 73.43.-f}

\abstract{
The conductivity $\sigma$ and resistivity $\rho$ tensors of the disordered Hofstadter model are mapped as functions of Fermi energy $E_F$ and temperature $T$ in the quantum critical regime of the plateau-insulator transition (PIT). The finite-size errors are eliminated by using the non-commutative Kubo-formula. The results reproduce all the key experimental characteristics of this transition in Integer Quantum Hall (IQHE) systems. In particular, the Quantized Hall Insulator (QHI) phase is detected and analyzed. The presently accepted characterization of the QHI phase in the quantum critical regime, based entirely on experimental data, is fully supported by our theoretical investigation.}

\begin{document}

\maketitle

The PIT between the Quantum Hall Liquid (QHL) phase $\sigma_{xy}=\frac{e^2}{h}$ and the insulating phase $\sigma_{xy}=0$ in IQHE systems is one of the most studied quantum transitions in condensed matter physics. The early experiments \cite{AlphenaarPRB1994gh,ShaharSSC1997re,PanPRB1997uy,HilkePRB1997tr,ShaharSSC1998bn}, including Ref.~\cite{HilkeNature1998fh} which provided the decisive evidence for the  QHI phase, probed only the classical regime. However, over the following years, the quantum critical regime at PIT was gradually conquered  \cite{SchaijkPRL2000vc,DunfordaPE2000gf,PonomarenkoPE2000yt,PruiskenSSC2006tr,VisserJPCS2006cu,LangPRB2007yg}. In the later experimental work \cite{VisserJPCS2006cu}, the critical behavior of the transport coefficients has been mapped with extraordinary experimental precision, enabling an accurate quantitative analysis of the phase diagram and scaling \cite{LangPRB2007yg}. This analysis \cite{LangPRB2007yg}, which shaped our current understanding of the PIT, revealed a large intrinsic finite-temperature scaling exponent and an intriguing evolution of the QHI phase in the extreme low-$T$ regime.  This phase appears in the insulator part of the transition, and is characterized by an exponentially small conductivity tensor but a finite quantized Hall resistivity $\rho_{xy}=\pm \frac{h}{e^2}$ (and diverging $\rho_{xx}$). In the quantum critical regime, the comprehensive analysis of Ref.~\cite{LangPRB2007yg} gave strong evidence that, while in the actual experimental range of temperatures ($0.1 < T < 1$ K) the QHI  occupies a substantial part of the phase diagram, an extrapolation to $T \rightarrow 0$, based on universal scaling arguments, showed that the QHI phase diminishes until it disappears. A confirmation of these conclusions by theoretical quantum simulations is highly desirable because here the sample's inhomogeneity can be directly eliminated. Such simulations must be necessarily carried at finite $T$'s and then extrapolated to $T=0$ using scaling arguments. Most of the existing theoretical studies have concentrated on the semiclassical regime \cite{ShaharSSC1997re,ShimshoniPRB1997nv,PryadkoPRL1999uf,ZulickeaPE2002bv,ShimshoniMPLB2004rt,LevyArxiv2010bv}, and the only few theoretical studies \cite{HuoPRL1993gf,ShengPRB1999pp,ShengPRB2000fr} performed in the quantum regime were carried at $T=0$ where the finite-size effects are inherently present.

In this work we report simulations for the disordered Hofstadter model \cite{HofstadterPRB1976km} based on the finite-temperature Kubo-formula for transport, as re-formulated by Schulz-Baldes and Bellissard \cite{BELLISSARD:1994xj,Schulz-Baldes:1998oq,Schulz-Baldes:1998vm} in the noncommutative geometry setting. This formalism enabled us to converge the simulations at $T$'s low enough to enter the quantum critical regime at PIT, where we were able to determine the scaling functions and extrapolate to $T=0$. Our results reproduce all qualitative and almost all quantitative experimental signatures of the PIT transition. Particularly, we report the detection of the QHI phase and a characterization of it that strongly supports the conclusions of Ref.~\cite{LangPRB2007yg}.

To understand the difficult simulation conditions and the solution provided by the non-commutative geometry formalism, we discuss first the physical regime where the QHI phase appears (based on \cite{LangPRB2007yg} and our own results). The $\sigma_{xy}=\frac{e^2}{h}$ and $\sigma_{xy}=0$ phases are separated by a phase boundary containing extended quantum states \cite{LevineNP1984tt,LevineNP1884hd,LevineNP1984uu,BELLISSARD:1994xj}. The diverging behavior of the localization length as $E_F$ approaches the phase boundary $E_c$ is characterized by the finite-size scaling exponent $\nu$: $\Lambda(E_F) \sim (E_F-E_c)^{-\nu}$. This law is accurate in a vicinity of $E_c$, or equivalently when $\Lambda(E_F)$ is large enough, lets us say: $\Lambda(E_F) > \bar{\Lambda}$, with $\bar{\Lambda}$ set by the sought accuracy. This $\bar{\Lambda}$ also sets the minimal system-size needed to observe the finite-size scaling of the conductance \cite{Abrahams:1979et,AndersonPRB1980cx}. When the $T=0$ picture is combined with the concept of the $T$-induced effective system-size, given by the Thouless length $L_{\mathrm{Th}}(T)\sim T^{-p/2}$ ($p=$ the dynamical exponent for dissipation) \cite{ThoulessPRL1977fj}, and with the single-parameter scaling hypothesis \cite{Abrahams:1979et,AndersonPRB1980cx}, one obtains the finite-$T$ scaling law \cite{PruiskenPRL1988cm}:
\begin{eqnarray}\label{TScaling}
\rho(E_F,T) =F\left((E_F-E_c)(T/T_0)^{-\kappa}\right),\ \kappa=p/2\nu.
\end{eqnarray}
Here, $\kappa$ is the finite-$T$ scaling exponent, $F$ is a system-dependent function and $T_0$ is a reference temperature. The scaling law of Eq.~\ref{TScaling}, which defines the quantum critical regime, goes into effect only if $L_{\mathrm{Th}}(T) > \bar{\Lambda}$. Furthermore, within the framework of Chalker-Coddington  network model \cite {ChalkerJPC1988}, the QHI phase can be observed only when the dephasing length is smaller than the typical size of the IQHE puddles \cite{PryadkoPRL1999uf,ShimshoniMPLB2004rt}. In a simulation like our, based on the Kubo formula with dissipation, this can be translated into $L_{\mathrm{Th}}(T) < \Lambda(E_F)$. Hence, our conclusion is that the QHI phase occurs when:
\begin{equation}\label{Domain}
\bar{\Lambda} < L_{\mathrm{Th}}(T) < \Lambda(E_F).
\end{equation} 
These conditions are represented in Fig.~1, and since $L_{\mathrm{Th}}(T) $ diverges as $T\rightarrow 0$, it can be seen why the QHI phase disappears at higher $T$'s, for $E_F$ far from $E_c$, or in the extreme limit $T \rightarrow 0$ (in line with \cite{LangPRB2007yg} and our own observations). 

\begin{figure}
\center
\includegraphics[height=4.5cm]{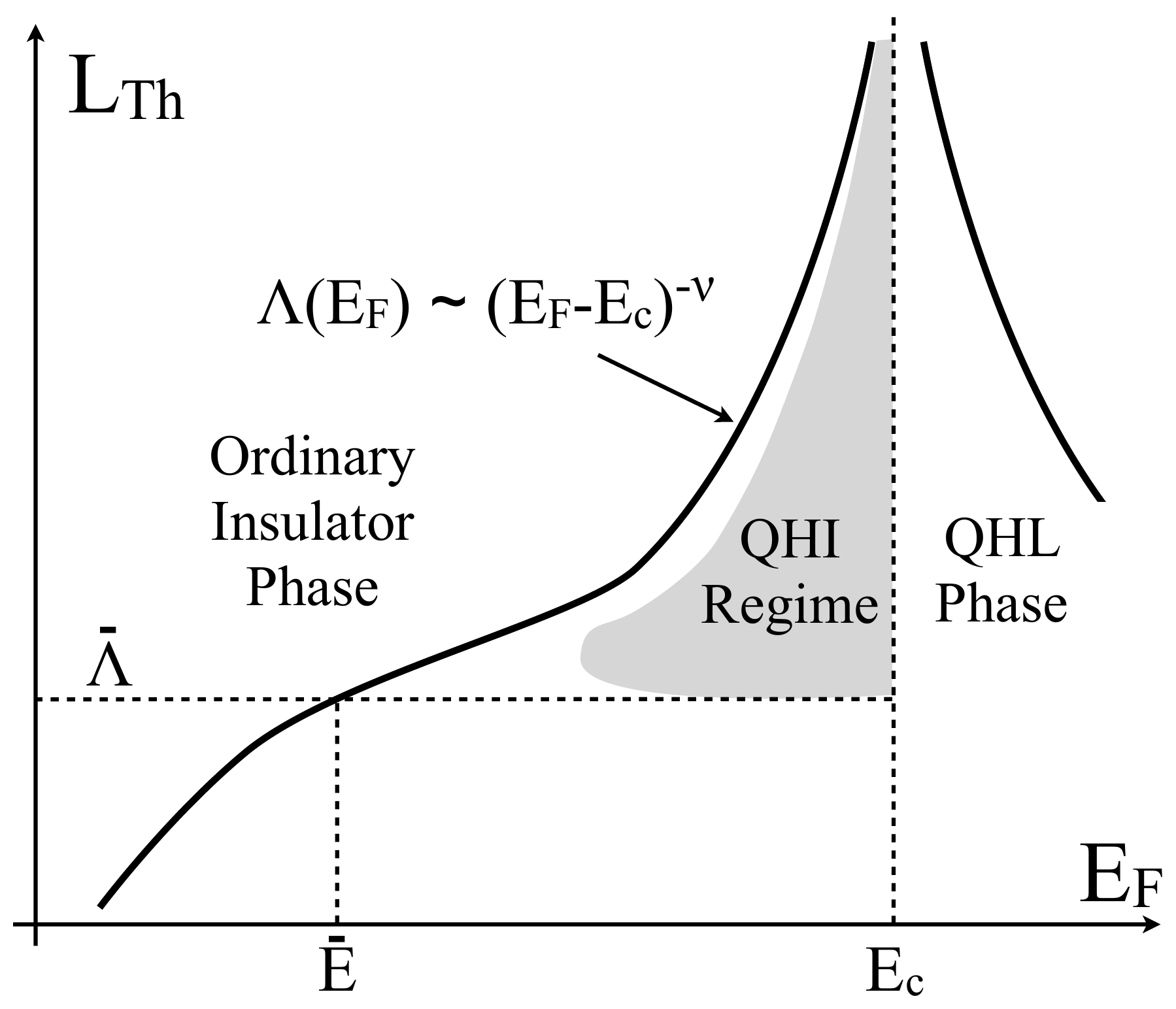}
\caption{The proposed phase diagram near PIT, drawn in the plane $(L_{\mathrm{Th}}, E_F)$. Note that $L_{\mathrm{Th}}$ depends monotonically on temperature hence the diagram can be re-drawn in the $(T,E_F)$ plane, but is more illuminating as presented.  The Quantized Hall Insulator is observed in the shaded region of the diagram. The boundary of the shaded region is not sharply defined, since the Quantized Hall Insulator morphs gradually into the ordinary Anderson insulator, without going through a phase transition.  The solid continuous lines represent the localization length $\Lambda(E_F)$ as function of Fermi energy. The shaded region ends before reaching the solid line (see text, especially Eq.~\ref{Domain}).}
\label{PhaseDiagram}
\end{figure}

This QHI regime is extremely difficult to simulate. First, one should note that the transport simulations simplify tremendously at $T=0$ or at high $T$'s. At $T=0$ and with no dissipation, the Hall conductance reduces to the Kubo-Chern formula \cite{ThoulessPRL1982vh,AvronPRL1983} which does not require the inversion of the Liouvillian (see Eq.~\ref{Kubo}). Furthermore, the diagonal conductivity can be obtained from the Landauer formula, which involves only the quantum states at $E_F$ \cite{ShengPRB2000fr}. In fact, the whole $\sigma$ can be obtained from a 4-terminal Landauer approach \cite{PryadkoPRL1999uf}. In the QHI regime, the entire energy spectrum has to be taken into account. The $T$'s are low so the details and the fine character of the energy spectrum are not washed away (as at high $T$'s) but instead they are the main factors determining the values of the transport coefficients. As such, the Liouvillian must be accurately inverted. But the main difficulty comes from the convergence of the numerical algorithms for Kubo-formula. The traditional implementations \cite{HouaritPPCM1991er,MandalPhysB1998as,RochePRB1999te,SteffenPRB2004bn} are known \cite{Xue2012fh} to converge only as an inverse power law in $L$ to the thermodynamic limit. Finding algorithms which converge exponentially fast was hindered by the fact that, for aperiodic systems, the Kubo-formula is usually presented as a formal limit as $L \rightarrow \infty$  [see for example Eq.~(3.385) and its finite-temperature version in Ref.~\cite{MahanBook2000bv}]. In contradistinction, the noncommutative Kubo-formula derived by Schulz-Baldes and Bellissard \cite{BELLISSARD:1994xj,Schulz-Baldes:1998oq,Schulz-Baldes:1998vm} provides an explicit thermodynamic limit which, together with the methods of noncommutative geometry used to derive it, enabled us in Ref.~\cite{ProdanAMRX2013bn} to establish (with mathematical rigor) a canonical finite-size approximation that converges exponentially fast in the thermodynamic limit.

In its full generality, the non-commutative Kubo-formula reads:
\begin{eqnarray}\label{Kubo}
\sigma_{ij}(T)=-\mathcal{T}\left ([X_i,H](\Gamma+\mathcal{L}_H)^{-1}[X_j,\Phi_{FD}(H)] \right ).
\end{eqnarray}
Here, $\mathcal{T}$ represents the trace over volume, $H$ is the disordered Hamiltonian, ${\bm X}$ is the position operator, $\mathcal{L}_H$ is the Liouvillian super-operator acting on operators as $\mathcal{L}_H(A) = \imath [A,H]$, $\Gamma$ is the dissipation super-operator which has a temperature dependence, and $\Phi_{FD}$ is the Fermi-Dirac distribution. Various models for the dissipation super-operator $\Gamma$, including the one implementing Mott's variable range hopping mechanism, and the physical regimes where these models are expected to apply are discussed in Refs.~\cite{Schulz-Baldes:1998oq,SpehnerJSP2001xr,BellissardLectNotesPhys2003cy,AndroulakisJSP2012je}. Here we will use a simplified version which assumes $\Gamma$ proportional to the identity $\Gamma = \frac{1}{\tau}$, with $\tau$ a $c$-number commonly referred to as the relaxation time. While this relaxation time approximation will be quite coarse in the middle of the QHE plateaus where the relevant electron wave functions are localized (in fact the extremely precise quantization of the plateaus can be explained only within Mott's variable range hopping picture \cite{BELLISSARD:1994xj}), the approximation is justified near the transition point where the electron wave functions are quite delocalized. An important remark is that the complexity of the calculations is the same regardless of the particular expression of $\Gamma$ being used. This is the case because any physically sound $\Gamma$ must commute with the Liouvillian and the latter is exactly diagonalized in our calculations. However, while the computational effort remains the same, the analysis and the interpretation of the results can become complicated when going beyond the relaxation time approximation. We leave such task for future investigations and here we only announce that numerical calculations with complex dissipation super-operators are now possible.

Regarding the canonical finite volume approximation $\sigma_{ij}^{(L)}(T,\tau)$ which is used in the present work,  Ref.~\cite{ProdanAMRX2013bn} established the error estimate:
\begin{eqnarray}
\left |\sigma_{ij}(T,\tau)-\sigma_{ij}^{(L)}(T,\tau) \right |<\textit{const.}\times e^{-\zeta(T,\tau)L},
\end{eqnarray}
which takes effect as soon as $L>L_{\mathrm{Th}}(T)$. The convergence rate $\zeta(T,\tau)$ was found \cite{ProdanAMRX2013bn} to be directly proportional with $T$ and inverse proportional with $\tau$, a conclusion that holds true regardless of the localized or delocalized nature of the energy spectrum at $E_F$. As such, our algorithm generates uniformly converged results with respect to $E_F$, which can be near or far from $E_c$. This is the strength of our approach.  A detailed discussion of the formalism and extensive convergence tests for the present model  at similar simulation conditions can be found in Ref.~\cite{ProdanAMRX2013bn}. A similar study for Quantum spin-Hall Insulators can be found in Ref.~\cite{Xue2012fh}.

\begin{figure}
\includegraphics[height=4.5cm]{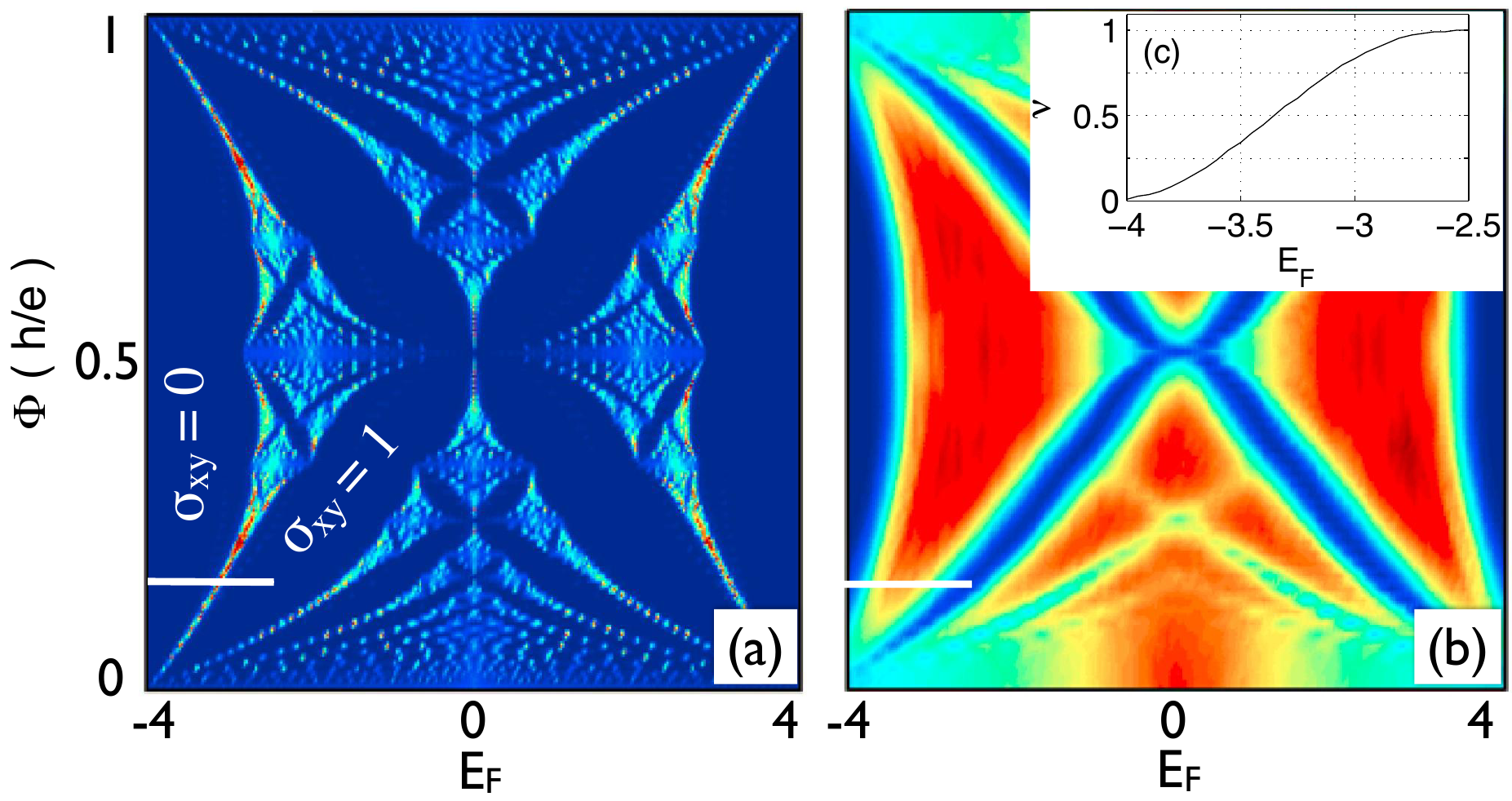}
\caption{ (Color online) The density of states of the model Eq.~\ref{Hofstadter} as function of the magnetic flux $\Phi$, for (a) $W=0$ and (b) $W=3$. The horizontal white line shows the range of energies and the $\Phi$ used in our simulations. The inset (c) shows the relationship between the filling factor $\tilde \nu$ and $E_F$.}
\label{DOS}
\end{figure}

We now present the details of the simulation and the results. We use the disordered Hofstadter model on a square lattice \cite{HofstadterPRB1976km}:
\begin{eqnarray}\label{Hofstadter}
H=-\sum_{\langle {\bm n},{\bm m} \rangle}{e^{-i \pi \Phi ({\bm n}\wedge {\bm m}) }c^\dagger_{\bm n} c_{\bm m}}
+W\sum_{\bm n} \omega_{\bm n} c^\dagger_{\bm n} c_{\bm n}.
\end{eqnarray}
Here, $\Phi$ is the magnetic flux per unit cell (in units $h/e$), ${\bm n}\wedge {\bm m}=n_1m_2-n_2m_1$, $c^\dagger_\mathbf{n}$ creates one electron at site ${\bm n}$ and $\langle {\bm n},{\bm m}\rangle$ means nearest neighboring sites. The $\omega$'s are independent random variables uniformly distributed in $[-\frac{1}{2},\frac{1}{2}]$. The simulations are performed on a $L=140$ lattice, with $W=3$ and $\Phi=22/L\approx 0.157$.  $\sigma$ is self-averaging, hence well converged results can be obtained even when averaging over a relatively small number ($= 53$ for the lowest $T$) of random configurations  \cite{ProdanAMRX2013bn}. Fig.~\ref{DOS} summarizes the spectral properties of the model and shows the energy range used in the simulations. The inset shows the variation of the filling factor $\tilde \nu$ with $E_F$, which is linear for a substantial energy range around $E_c$. As such, we can represent our data as function of either $E_F$ or $\tilde \nu$. Here we choose $E_F$ and in the supplemental material we show the data as function of $\nu$.

\begin{figure}
\includegraphics[height=6.5cm]{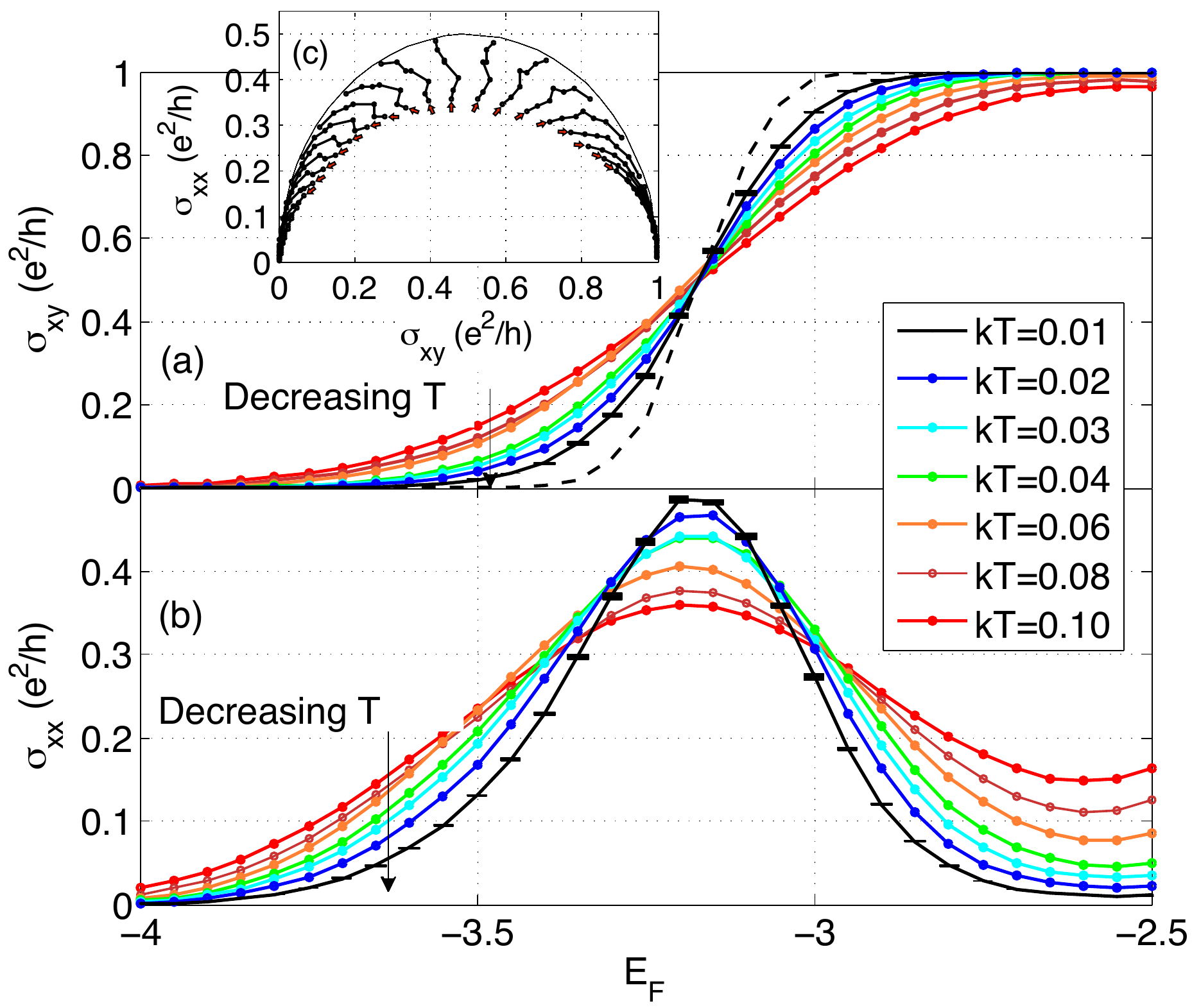}
\caption{ (Color online) The simulated (a) $\sigma_{xy}$ and (b) $\sigma_{xx}$, as functions of $E_F$ at different $T$'s. The dashed line in panel (a) represents the Chern number. The marker-sizes (black rectangle) corresponding to $kT = 0.01$ reflect the actual statistical errors. The inset (c) shows the flow of $\sigma$ with $T$.}
\label{SigmaVsEf}
\end{figure}

Fig.~\ref{SigmaVsEf} reports the simulated $\sigma$ as function of $E_F$, at $k T=1/\tau=0.01, 0.02, 0.03, 0.04, 0.06, 0.08$ and $0.10$ ($k$ = Boltzmann constant). Note that we are fixing $p=1$ by making these choices, but this does not put a limitation on our conclusions because any other choice for $p$ would only rescale the horizontal axis of our graphs (provided $1/\tau$ varies within the same bounds, which can be enforced by a proper choice of the constant $\alpha$ in $1/\tau = \alpha T^p$).  We have also included a calculation of the Chern number ($=\sigma_{xy}$ at $T=0$), computed with the numerical algorithm detailed in \cite{ProdanPRL2010ew,ProdanJPhysA2011xk}. Since this algorithm converges exponentially fast in the thermodynamic limit once $L>\Lambda(E_F)$, a precise quantization of the Chern number implies $L>\Lambda(E_F)$. Recall that the convergence of $\sigma$ at finite $T$'s is conditioned by $L>L_{\mathrm{Th}}(T)$, which is the case in Fig.~\ref{SigmaVsEf}. Furthermore, the statistical errors are smaller than the marker-sizes. In panel (a) we see $\sigma_{xy}$ transitioning from $e^2/h$ to 0, and the transition becoming sharper as $T \rightarrow 0$. To a high degree, all $\sigma_{xy}$-curves intersect each other at one point, the critical point $E_c$. The slope of the curves at $E_c$ obey the well known scaling with T \cite{PruiskenPRL1988cm}. In panel (b), $\sigma_{xx}$ displays an insulating behavior for most energies, with $\sigma_{xx}$ decreasing as $T \rightarrow 0$, except for a small region around $E_c$ where  $\sigma_{xx}$ increases as $T \rightarrow 0$, signaling the presence of extended quantum states. The data suggests that this region subsides to one point and that the maximum value of $\sigma_{xx}$ saturates at $\frac{1}{2}\frac{e^2}{h}$ as $T \rightarrow 0$, in line with the accepted theoretical picture of PIT \cite{PruiskenPRL1988cm}. The inset in Fig.~\ref{SigmaVsEf} shows the flow of $\sigma$ with $T$. There is quite a substantial dependence on $T$, but $\sigma$ ultimately flows to a separatrix which is shaped as a perfect semi-circle. An unstable fixed-point at $\sigma_{xy}=\frac{1}{2}\frac{e^2}{h}$ can be clearly seen, in agreement with the renormalization theory of the IQHE \cite{PruiskenPRL1988cm,PruiskenEL1995tr}. In the semi-classical regime, the semicircle law follows from general theoretical arguments in a ``two-phase model" \cite{DykhnePRB1994re}, and is equivalent to the quantization of $\rho_{xy}$ at $\frac{h}{e^2}$ on both sides of the transition \cite{HilkeNature1998fh}. In the quantum critical regime, the experimental T-flow of $\sigma$ was reported in Ref.~\cite{SchaijkPRL2000vc}, and it looks very similar to the flow of $\sigma$ in our simulations. This indicates already that the QHI phase is present in our simulations.

\begin{figure}
\includegraphics[height=5.5cm]{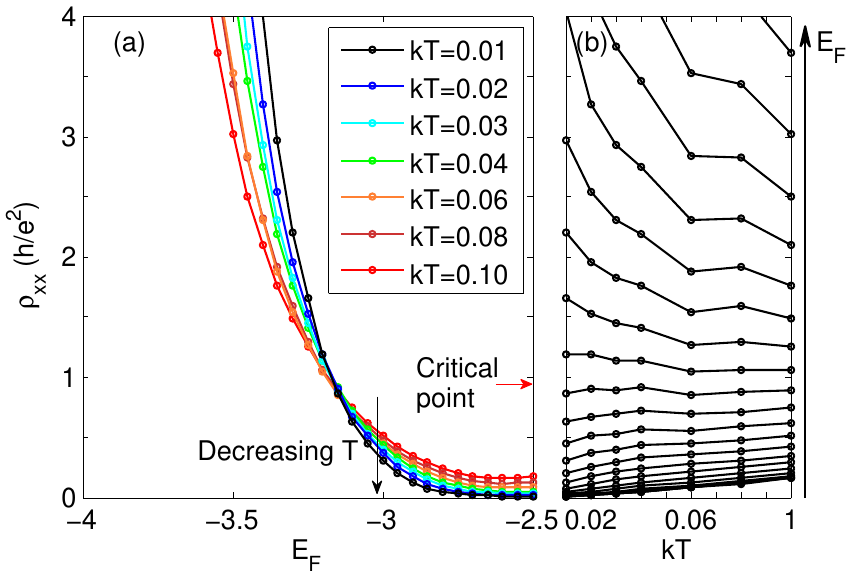}
\caption{ (Color online) (a) $\rho_{xx}$ as function of $E_F$, at different $T$'s. (b) $\rho_{xx}$ as function of $T$ for various $E_F$ values. The arrow (red) indicates the plateau-insulator transition.}
\label{RhoXX}
\end{figure}

The existence of a single critical point, as opposed to a line of critical points, is best revealed in Fig.~\ref{RhoXX}, which reports $\rho_{xx}=\sigma_{xx}/(\sigma_{xx}^2+\sigma_{xy}^2$) as function of $E_F$ (panel a), and as function of $kT$ (panel b). Here, one can see $\rho_{xx} \rightarrow \infty$ in the QHI side, and $\rho_{xx} \rightarrow 0$ in the QHL side. While the latter behavior is usually attributed to a metallic phase, in the present context it is due to the fact that $\sigma_{xy}\neq 0$. These opposite behaviors make the $\rho_{xx}$-curves intersect each other, very much like $\sigma_{xy}$-curves do in Fig.~\ref{SigmaVsEf}(a). To a high degree, all the curves (even those at higher $T$-s) intersect at a single critical point, exactly as the experimental data showed (see Fig.~1 in Refs.~\cite{HilkeNature1998fh}, \cite{SchaijkPRL2000vc} and \cite{VisserJPCS2006cu}). Note that all PIT experiments were done at $T$'s larger then or comparable to $0.1$ K, quite far from the extreme low-$T$ regimes probed by $T=0$ theoretical simulations \cite{HuoPRL1993gf}. As already demonstrated experimentally \cite{HilkePRB1997tr}, the coordinates of the critical point can be extracted with great precision if $\rho_{xx}$ is plotted as function of $kT$, as in Fig.~\ref{RhoXX}(b). Here one can see that, as $T \rightarrow 0$, the $\rho_{xx}$-lines curve  downwards/upwards below/above a sharp critical point $E_c \approx -3.15$. The critical value $\rho_{xx}^c$ at PIT is virtually equal to $h/e^2$, and from Fig.~\ref{SigmaVsEf}(a) we estimate $\sigma_{xy}^c$ to be virtually equal to $\frac{1}{2}\frac{e^2}{h}$, exactly as in experiments \cite{HilkeNature1998fh,SchaijkPRL2000vc,VisserJPCS2006cu}. 

\begin{figure}
\includegraphics[height=5.5cm]{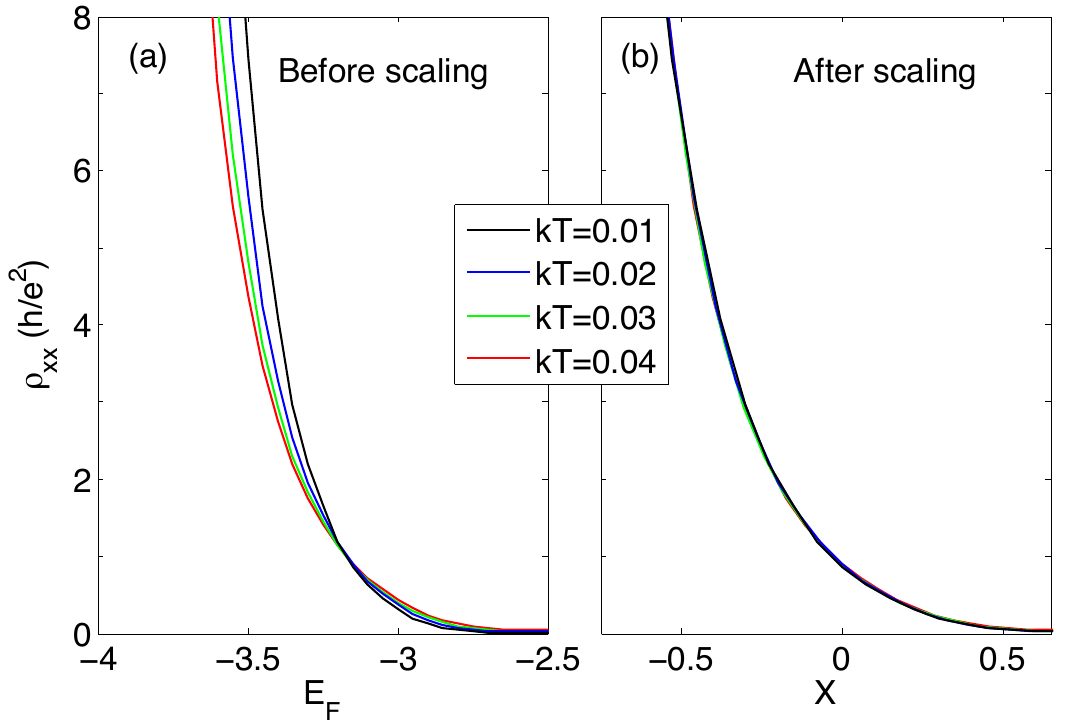}
\caption{ (Color online) $\rho_{xx}$ data (a) before and (b) after the rescaling: $E_F \rightarrow X=(E_F-E_c)\left ( \frac{T}{T_0}\right ) ^{-\kappa}$, with $E_c=-3.15$, $kT_0=0.08$ and $\kappa=0.197$.}
\label{Scaling}
\end{figure}

In Fig.~\ref{Scaling} we demonstrate that we entered the quantum critical regime with the four lowest $T$'s. Indeed, upon a single-parameter rescaling of the energy axis, $E_F \rightarrow X=(E_F-E_c)\left ( \frac{T}{T_0}\right ) ^{-\kappa}$, the $\rho_{xx}$-curves collapse almost perfectly on top of each other. The best overlap was obtained for $\kappa=0.197 \pm 0.004$, which compares extremely well with the value $\kappa = 0.194 \pm 0.002$ obtained from $\kappa = p/2\nu$ with the presently accepted (average) theoretical value $\nu=2.58 \pm 0.03$ \cite{SlevinPRB2009tr,KramerIJMPB2010wk,ObusePRB2010fj,FulgaPRB2011cv,DahlhausPRB2011bn,AmadoPRL2011fj,SlevinIJMP2012gh,ObusePRL2012ls} and $p=1$ like in our simulations. Experimentally, $\kappa$ was consistently found \cite{SchaijkPRL2000vc,DunfordaPE2000gf,PonomarenkoPE2000yt,PruiskenSSC2006tr,VisserJPCS2006cu,LangPRB2007yg} to be much larger at PIT ($\kappa\approx 0.58$) than at the plateau-plateau transitions ($\kappa\approx 0.42$ \cite{WanliPRL2009gf}). Explaining this discrepancy remains an open problem \cite{VisserJPCS2006cu}. A possible explanation could have been the breakdown of the relation $\kappa = p/2\nu$ at PIT, due to the presence of the QHI phase. As we shall see, the QHI phase is already present in our simulations, yet we observe that the law $\kappa = p/2\nu$ is very accurate at PIT.

\begin{figure}
\includegraphics[height=7cm]{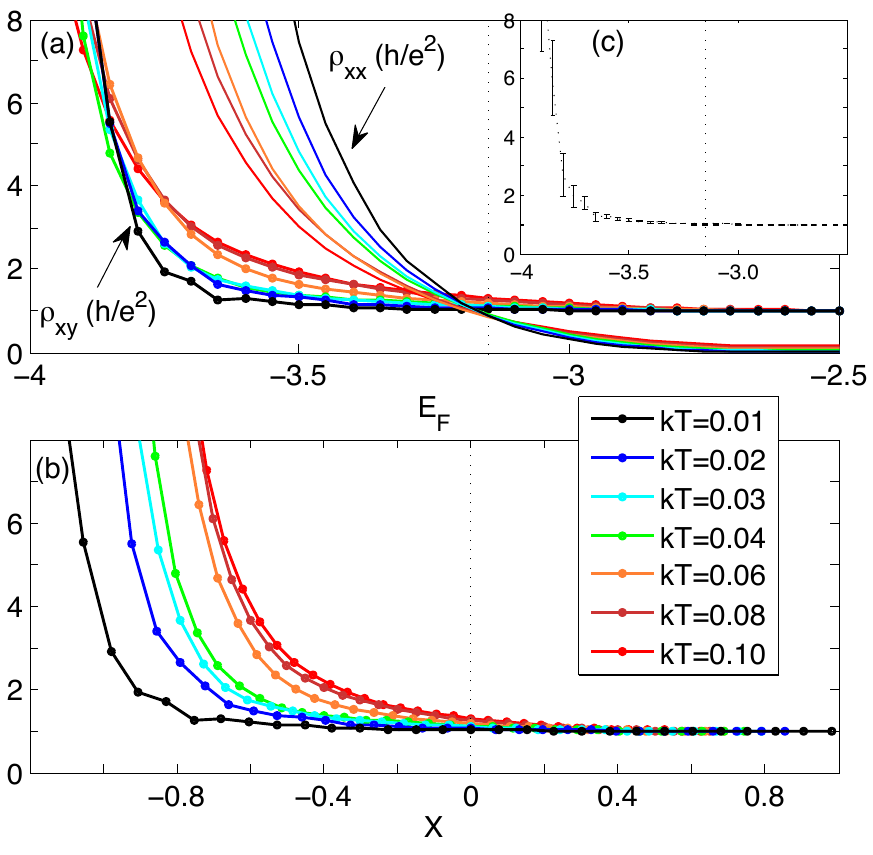}
\caption{ (Color online) (a) $\rho_{xy}$ and $\rho_{xx}$ as functions of $E_F$, at different $T$'s. (b) $\rho_{xy}$ represented as function of $X$. The inset (c) shows the statistical error for $\rho_{xy}$ at $kT=0.01$.}
\label{QHI}
\end{figure}

We now come to the main finding of our work. Fig.~\ref{QHI}(a) reports $\rho_{xy}$ as a function of $E_F$ at various $T$'s. As one can see, the $\rho_{xy}$-curves flatten out and converge to the line $\rho_{xy}=\frac{h}{e^2}$ as $T \rightarrow 0$, well beyond the critical point in the insulator side of PIT. Three simulated quantized $\rho_{xy}$-values occur at $E_F$'s where $\sigma_{xy}|_{T=0}=0.0005, \ 0.004$, and $0.01$, indicating that $L>\Lambda(E_F)$ for these data points. This rules out any possible artifacts and we can conclude beyond any doubt that the QHI has been detected by our simulations. Furthermore, Fig.~\ref{QHI}(a) looks similar to the representation of the experimental data reported in Fig.~3 of Ref.~\cite{LangPRB2007yg}. We can see that, further away from $E_c$, the behavior of $\rho_{xy}$ with $T$ is not monotonic, and there seems to be a threshold $T$  below which $\rho_{xy}$ is reversing its trend and start diverging as $T$ is decreased. In Ref.~\cite{LangPRB2007yg} it was suggested that the scaling variable $X$ is more appropriate for representing the phase diagram, so in Fig.~\ref{QHI}(b) we show $\rho_{xy}$ as function of $X$. The similarity with Fig.~2 of Ref.~\cite{LangPRB2007yg} is evident. Hence our data seems to be in agreement with the findings in Ref.~\cite{LangPRB2007yg} and to support the conclusion that the QHI phase disappear at $T=0$ if the phase diagram is represented as function of $E_F$ (or $\tilde \nu$), but it fills the entire insulator space if represented as function of $X$. To fully resolve the extreme limit $T \rightarrow 0$, we fitted the rescaled curves reported in Fig.~\ref{Scaling}(b) and Fig.~\ref{QHI}(b) with the same analytic expressions used in Ref.~\cite{LangPRB2007yg}:
\begin{equation}
\rho_{xx}(X)=e^{-X-\gamma X^3}, \ \ \rho_{xy}=1+(T/T_1)^{y}\rho_{xx}(X).
\end{equation}
The overlap between the best fit and the data is almost perfect. Among the fitting parameters $\gamma$, $T_1$ and $y$, the most important one is $y$ because a positive value supports the scenario at $T=0$ described above, and a negative one does not. We found $y=0.165$, a substantial positive value. 

In conclusion, simulations based on the noncommutative Kubo-formula enabled us to enter the quantum critical regime at PIT and to detect and characterize the QHI phase. Our theoretical analysis fully supports the presently accepted characterization of the QHI phase in the quantum critical regime.

\acknowledgments We acknowledge stimulating discussion with Yigal Meir. This work was supported by the U.S. NSF grants DMS 1066045, DMR-1056168, NSFC under grants No.
11204065 and RFDPHE-China under Grant No. 20101303120005.

\bibliographystyle{eplbib}
\bibliography{../../../../TopologicalInsulators}

\end{document}